# Development of MicroTCA.4 based remote DAQ system for KSTAR

Giil Kwon, Woong-ryol Lee, Taegu Lee, Kihyun Kim, Jaesic Hong

*Abstract*— **To standardize and simplify the control system at Korea Superconducting Tokamak Advanced Research (KSTAR), we develop 10G Ethernet based remote DAQ system. By separating the DAQ system and the host, the structure of the control system can be made more flexible. We have developed a DAQ module with a 10G Ethernet interface based on a MicroTCA.4 system designed to control devices in real time on a remote server via 10GE. To connect proposed device and host, we use real time network based on UDP multicast atop 10GbE cut-through packet switching infrastructure. This system is implemented using Zynq based MicroTCA.4 board, matched RTM board that has analogue input/output interface and power supply system. By using remote DAQ system, multiple host server can subscribe the DAQ data without additional computational cost in real time. This system will be applied to control fueling system at KSTAR Tokamak.**

*Index Terms*— **10G Ethernet, Data acquisition, MicroTCA.4,**

## I. INTRODUCTION

THE KSTAR is a Superconducting Tokamak at Korea. KSTAR consists of various kinds of equipment and systems that control it. However, these various control systems are difficult to maintain. To solve this problem, KSTAR has developed standardized control devices such as KMCU[3]. Some KSTAR devices require only a small number of high-speed(>1kHz) AI channels and AO channels (Such as Fueling system, Neutral Beam Injection system). The use of DAQ equipment with MicroTCA.4 crates in such equipment is not suitable in terms of cost or complexity. To simplify the control system and develop a flexible control structure, we have developed a DAQ module with a 10G Ethernet interface based on a MicroTCA.4 system designed to control devices in real time on a remote server via 10G Ethernet. This system is a more flexible and compacter than existing DAQ system. This is because the host and the DAQ module can be separated and multiple hosts can be connected to the DAQ module. There are some MicroTCA.4 device that can connect to host server with Optical link [1][2]. These devices provide the optical PCIe uplink to remote host server. However, these devices are expensive and can only connect one host. Proposed system is based on 10G Ethernet.

This work was supported in part by Korean Ministry of Science, ICT and Future Planning. The work supported by KSTAR team.

Giil Kwon, Woong-ryol Lee, Taegu Lee, Jaesic Hong is with the National Fusion Research Institute, Daejeon, Korea (telephone: +82-42-879-5238, e-mail: giilkwon@nfri.re.kr).

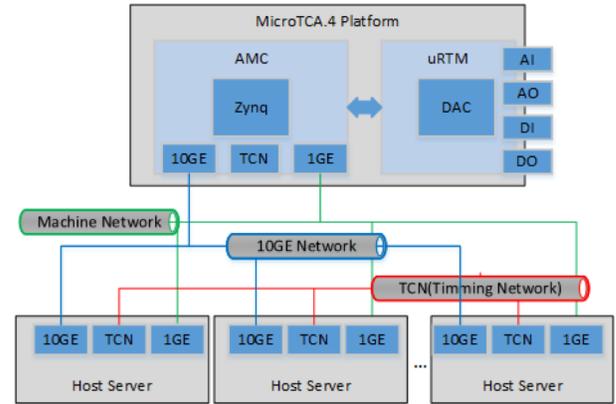

Fig. 1. Basic Concept and Architecture of System.

This system and host are connected with 10G Ethernet. Therefore, existing 10G Ethernet network infrastructure can be used to connect device with host server. Our system support 10G Ethernet UDP multicast. Multiple host server can subscribe the packet without additional computational cost. And can publish the packet by using dedicated topic.

## II. DESIGN AND DEVELOPMENT

The architecture of the proposed system is described in Figure. 1. Our system DAQ has AI, AO, DI and DO port. Host server and our system connected to 1G ethernet based machine network and 10G Ethernet. We send and receive parameter data

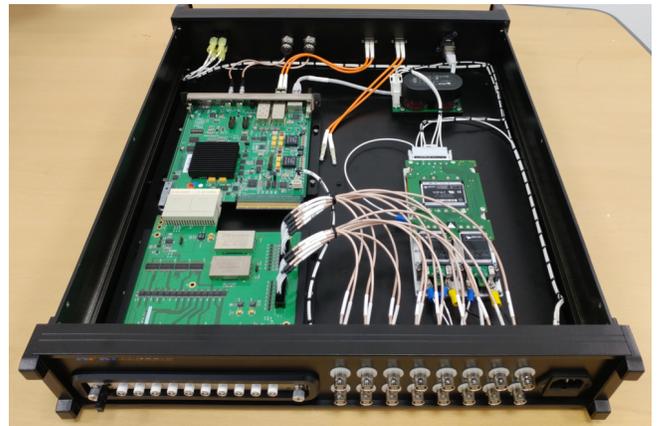

Fig. 2. Manufactured MicroTCA.4 based remote DAQ.

Kihyun Kim is with the Seedcore Ltd., Daejeon, Korea (telephone: +82-42-867-6501, e-mail: khkim5@seedcore.co.kr).

via 1G Ethernet network. We transmit large size data through 10G Ethernet network. Manufactured device is illustrated in Figure. 2. This system consist of homemade case, power, MicroTCA.4 AMC (Advanced Mezzanine Cards) board and RTM (Rear Transition Module) board. This system does not contain MCH and Backplane of MicroTCA.4 to make compact system.

*A. Hardware Design*

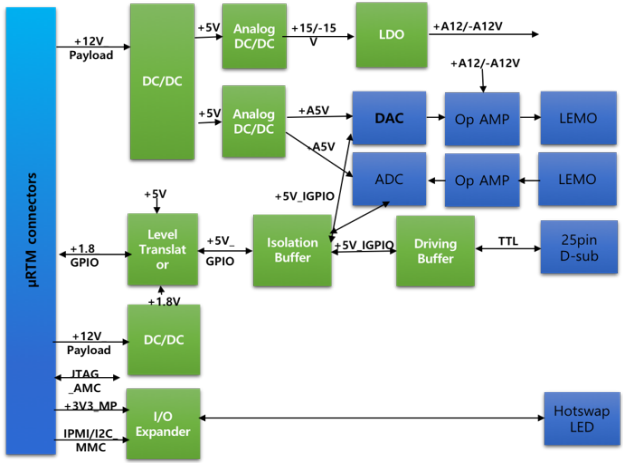

Fig. 3. Block diagram of RTM board.

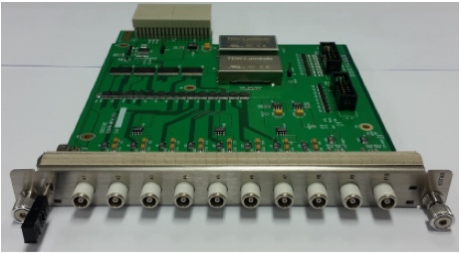

Fig. 4. Manufactured μRTM board.

This system is implemented using Zynq based MicroTCA.4 AMC board, matched RTM board that has analogue/digital input/output interface and linear power supply system. Linear power supply system has low noise. To develop this system, we use existing Zynq based MicroTCA.4 AMC board, KMCU. KMCU was developed to take advantage of the modular structure and the flexible reconfiguration capability of the MicroTCA.4 standard. We also design and develop RTM board with analog Data Acquisition (DAQ) modules. RTM board has 2 channels ADC (18 bit resolution, 1MSPS/channel), 8 channels DAC (16 bit resolution, 1MSPS/channel), 8 channels (TTL Input), 8 channels(TTL Output).

Figure 3. Describes the block diagram of this analogue interface board. And Fig. 4. Show use the manufactured RTM board.

*B. Software Design*

To connect proposed device and host, we use Synchronous Data bus Network (SDN) protocol. SDN is ITER real time network based on UDP multicast atop 10GbE cut-through packet switching infrastructure. SDN has 4 μsec latency with 64bytes payload. Figure 5. Show us the block diagram of FPGA program. This module parse UDP packet from 10GE module and transmit this data to DAQ module to generate analogue output. Currently, Logics for AI/DI/DO module is not implemented. Figure 6 describe the software architecture of host program. Host side program is implemented using TAC-engine [5]. TAC-engine is real-time framework which consist of multi-thread and support SDN and EPICS. Host side program synchronized with KSTAR with ITER Time Communication Network (TCN)[1].

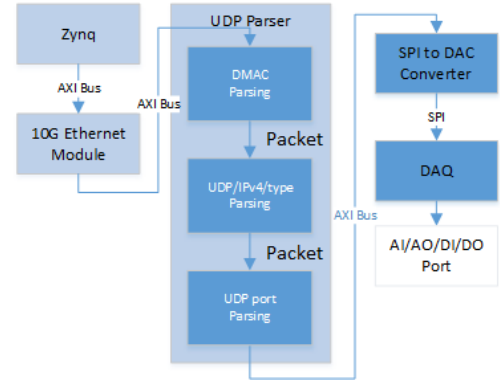

Fig. 5. Block diagram of FPGA program.

The first target system will be tokamak fueling system. For plasma generation and density control, gas is controlled by receiving real-time density control signal from the Plasma Control System (PCS). Remote DAQ system will controls supplied gas volume by receiving the SDN packet that contain real-time command from the PCS and controlling Piezo Electric Valve (PEV). This control cycle will be 2 kHz.

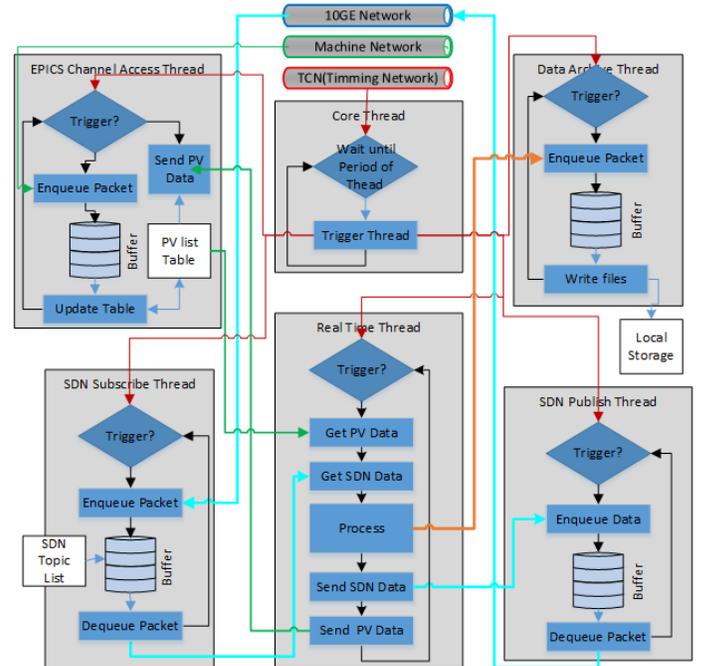

Fig. 6. Software Architecture of Real Time Framework, TAC-engine.

*C. Real-Time Tuning*

To get deterministic latency of SDN, we did real time tuning to host server. Host server OS is Centos 7.5 and Real Time kernel is used to get real time performance. And we applied real time profile to the system. We turn off the hyper thread and power saving option from bios. We also isolate 1-3 CPUs for our application. Only CPU 0 is used for general purposes. And CPU1-3 is used for our real time application.

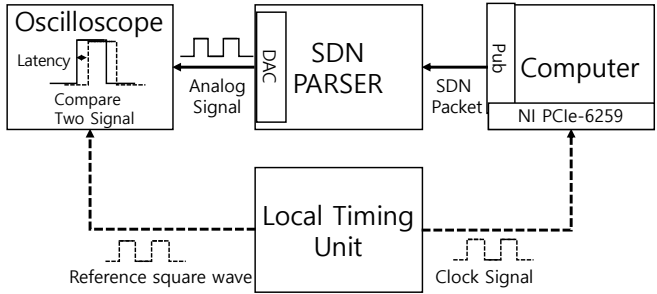

Fig. 7. Test environment for measuring total time it took for a packet to be transmitted and generating a signal.

## III. EXPERIMENT

To evaluate the performance of SDN-parser system, we did two latency tests. First, we measured the total time it took for a packet to be transmitted and generating a signal. And Second, we measured the time it takes to receive a packet and generate an analog signal from the device.

The test environment is as followed. In order to accurately measure the time it took for a packet to be transmitted and generating a signal, we use Local Timing Unit (LTU). KSTAR uses LTU to synchronize the system with each other. LTU can generate multiple clock signal at the same time. The time difference between theses signal is less than 200 picoseconds [6]. We compare clock signal from LTU and the signal from SDN-Parser to get the latency of system. To acquire the clock signal from LTU, we installed the NI PCIe-6259 DAQ card and get clock signal from the DAQ. The test program operates in synchronization with the clock signal. As test program got clock signal, the program generates SDN packet and send this packet to the device. The clock signal frequency is 10kHz and the test program also work as 10kHz.

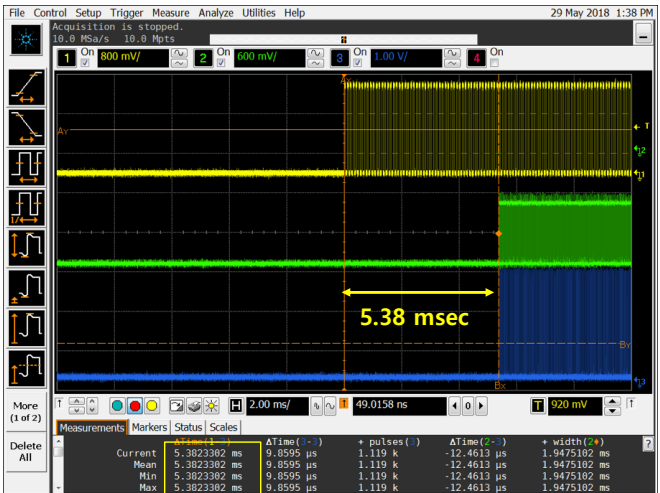

Fig. 8. Test result for measuring total time it took the server to send a packet and generate an analog signal on the device.

The SDN packet consist of 48bytes Header and 36bytes payload. This payload contains one channel signal value. Each packet only can contain one channel signal.

Fig. 8. Show us the result of test. Yellow line indicates us the clock signal from LTU, green line represents the generated analog signal from SDN-Parser and blue line indicates the packet acceptance signal which is generated as the SDN-Parser receive the packet from the server. The time difference between yellow line and green line represents the total latency of the system. This figure shows us that the server takes 5.38 millisecond to send a packet and generate a signal from the device.

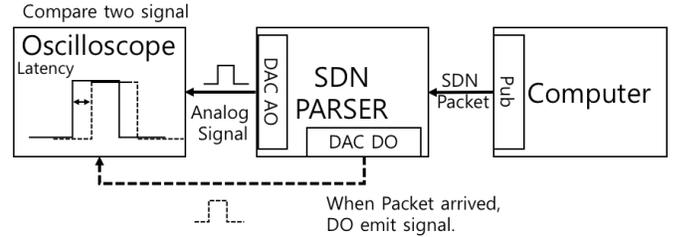

Fig. 9. Test environment for measuring the time it takes to receive a packet and generate an analog signal from the device.

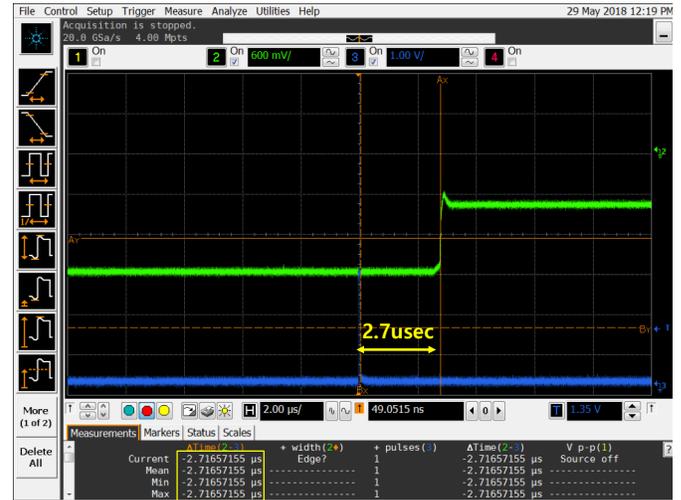

Fig. 10. Test environment for measuring the time it takes to receive a packet and generate an analog signal from the device.

In order to accurately qualify the SDN-Parser's performance, we measured the time it takes to receive a packet and generate an analog signal from the device. This value indicates the latency of the SDN-Parser.

As Fig.9. show us, SDN-Parser generates digital signal when the system received the packet from host server. We compare this signal and generated analog signal from SDN-Parser to measure the latency of the SDN-Parser.

In Fig. 10, Green line represents the generated analog signal from SDN-Parser and blue line indicates the packet acceptance signal. SDN-Parser emit digital signal when the system receive packet from the server. The SDN-Parser parse the packet and generates analog signal. The time difference between these two signals is 2.7 μsec. Considering that SDN-Parser has a latency of 2.7 μsec, it takes 5.379 msec for the server to send packets according to the clock. To reduce this time, we will do more tuning to host server.

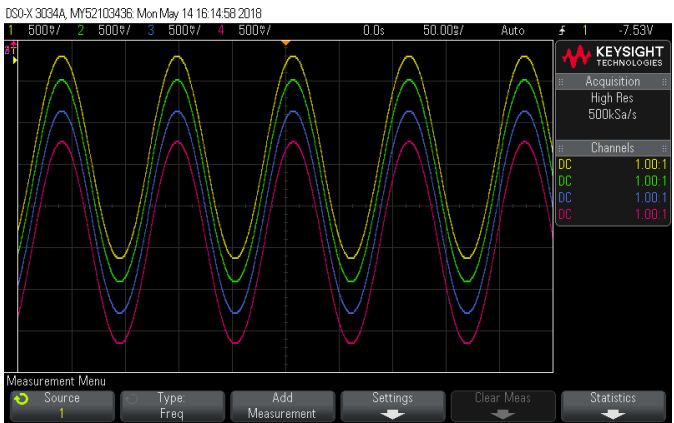

Fig. 11. Generated 4 channel analog signal captured by oscilloscope. SDN-Parser can generate 8 channel analog signal in parallel.

## IV. CONCLUSION

We have developed a DAQ module with a 10G Ethernet interface based on a MicroTCA.4 system designed to control devices in real time on a remote server via 10G Ethernet. This system is implemented using Zynq based MTCA board and matched DAQ RTM board. By using the proposed system, multiple host servers can be connected to the proposed system for real-time control from remote place. This system is suitable for environments where host server and DAQ equipment should be separated.